\documentclass[12pt]{iopart}

\usepackage{iopams}
\usepackage{graphicx}
\begin{document}

\title[Autoionization dynamics and sub-cycle beating in electronic molecular wave packets]{Observation of autoionization dynamics and sub-cycle quantum beating in electronic molecular wave packets}

\author{M.~Reduzzi$^{1,3}$, W.-C.~Chu$^{2}$, C.~Feng$^{1}$ A.~Dubrouil$^{1}$, J.~Hummert$^{1}$, F.~Calegari$^{3}$,
F.~Frassetto$^{4}$, L.~Poletto$^{4}$, O.~Kornilov$^{5}$, M.~Nisoli$^{1,3}$, C.-D.~Lin$^{2}$, G.~Sansone$^{1,3}$}

\address{(1) Dipartimento di Fisica, Politecnico Piazza Leonardo da Vinci 32, 20133 Milano Italy\\
(2) Physics Department, Kansas State University, Manhattan, Kansas 66506, U.S.A.\\
(3) Institute of Photonics and Nanotechnologies, CNR Politecnico, Piazza Leonardo da Vinci 32, 20133 Milano Italy\\
(4) Institute of Photonics and Nanotechnologies, CNR via Trasea 7, 35131 Padova Italy\\
(5) Max-Born-Institut, Max Born Strasse 2A, D-12489 Berlin\\
email: giuseppe.sansone@polimi.it}
\begin{abstract}
The coherent interaction with ultrashort light pulses is a powerful strategy for monitoring and controlling the dynamics of wave packets in
all states of matter. As light presents an oscillation period of a few femtoseconds ($T=2.6$~fs in
the near infrared spectral range), the fundamental light-matter interaction occurs on the sub-cycle timescale, i.e. in a few hundred attoseconds. In this work, we resolve the dynamics of autoionizing states on the femtosecond timescale and observe the sub-cycle evolution of a coherent electronic wave packet in a
diatomic molecule, exploiting a tunable ultrashort extreme ultraviolet pulse and a synchronized infrared
field. The experimental observations are based on measuring the variations of the extreme ultraviolet
radiation transmitted through the molecular gas. The different mechanisms contributing to the wave packet dynamics are investigated through theoretical simulations and a simple three level model. The method is general and can be extended to the investigation of more
complex systems.
\end{abstract}

\maketitle

\section{Introduction}
Controlling the motion of electrons in a molecule during the unfolding of a chemical process is
one of the main goals of attosecond science~\cite{RMP-Krausz-2009, NATPHYS-Corkum-2007, PE-Nisoli-2009}. As chemical bonds
are associated with the (multi-)electronic wavefunction distribution, the possibility to
observe and steer the electronic motion in real-time with field-controlled pulses opens new
perspectives in the manipulation of chemical reactivity. This novel approach for
molecular quantum control is at the heart of the expanding field of attosecond chemistry
and it is attracting an increasing number of theoretical and experimental
efforts~\cite{CHEMPHYSLETT-Lepine-2013, SCIENCE-Kapteyn-2007}.
Electron molecular dynamics is initiated with the creation of an electronic
wave packet. Using an extreme ultraviolet (XUV) attosecond pulse, wave packets comprising bound and continuum states
can be excited \cite{PRL-Mauritsson-2010}. The subsequent ultrafast dynamics can be observed by processes such as charge
rearrangement on a timescale of few tens of attoseconds~\cite{PRL-Breidbach-2005}, or by the
oscillation of the charge cloud over the entire molecular structure on a timescale of few
femtoseconds~\cite{PNAS-Remacle-2006}. For longer timescales, the nuclear motion comes into play
calling for the coupled description of the electronic and nuclear degrees of
freedom~\cite{JCP-Cederbaum-2008}.

In 2010, the first experimental demonstration of the
attosecond control of the electronic motion in a dissociating molecule (hydrogen and deuterium)
was reported, measuring the emission direction of the ionic fragments
resulting from molecular dissociation~\cite{NAT-Sansone-2010}.
Photoelectron and photoion spectroscopy, however, are based
on the ionization (and dissociation) of the molecular system, implying photoionization as the
initial trigger of the molecular dynamics. The creation and manipulation of bound electronic
states in the neutral molecule are of extreme interest for chemical
reactivity control, pointing out the need of a general experimental technique to access the dynamics of
bound states.

In this context, it is important to point out that the excitation of an electronic wave packet corresponds to the creation of a time-dependent electronic charge distribution, and therefore to a dipole moment $d(t)$. An initial study of the
dipole in neutral molecules induced by an intense laser pulse and measured by a time-delayed
ionizing XUV pulse was reported recently~\cite{PRL-Neidel-2013}. A more direct method for gaining insight in the
dynamics of the system is the spectral characterization of the induced radiation.
In this picture, the initial attosecond pulse creates a coherent population of states, triggering
the polarization of the medium. The dynamics occurring after the excitation pulse is
encoded in the time-dependent polarization and, therefore, in the spectral amplitude and phase
of the radiation transmitted through the system, giving access to the
evolution of the electronic states. This technique, usually named attosecond transient
absorption spectroscopy, has been applied for the investigation of noble gas
atoms, gaining access to the evolution of singly excited states in helium~\cite{PRA-Chen-2012,
SR-Chini-2013, PRL-Holler-2011, NJP-Lucchini-2013} and neon~\cite{PRA-Wang-2013}, and doubly excited states in argon~\cite{PRL-Wang-2010}
and helium~\cite{CP-Loh-2008, NATURE-Ott-2014, SCIENCE-Ott-2013}.

Attosecond broadband wave packets spanning across the ionization threshold have also been studied in time-delayed
photoelectron measurement~\cite{PRL-Mauritsson-2010}; however, the
bound state part of the wave packet could be revealed only indirectly in the electron measurement.

In this work, we apply attosecond transient absorption to a
molecular system, demonstrating the creation of a complex electronic wave packet composed of
bound excited states and autoionizing resonances. Using a synchronized
infrared (IR), carrier-envelope-phase (CEP) stable field, the evolution of the wave packet can be
observed in time. By measuring the broadband XUV absorption cross
section vs the IR time delay, we extract information on the autoionization dynamics and on the population redistributions in
the different final states of the molecule.
\section{Three-level model}
\begin{figure}[htb]
\centering\includegraphics[width=14cm]{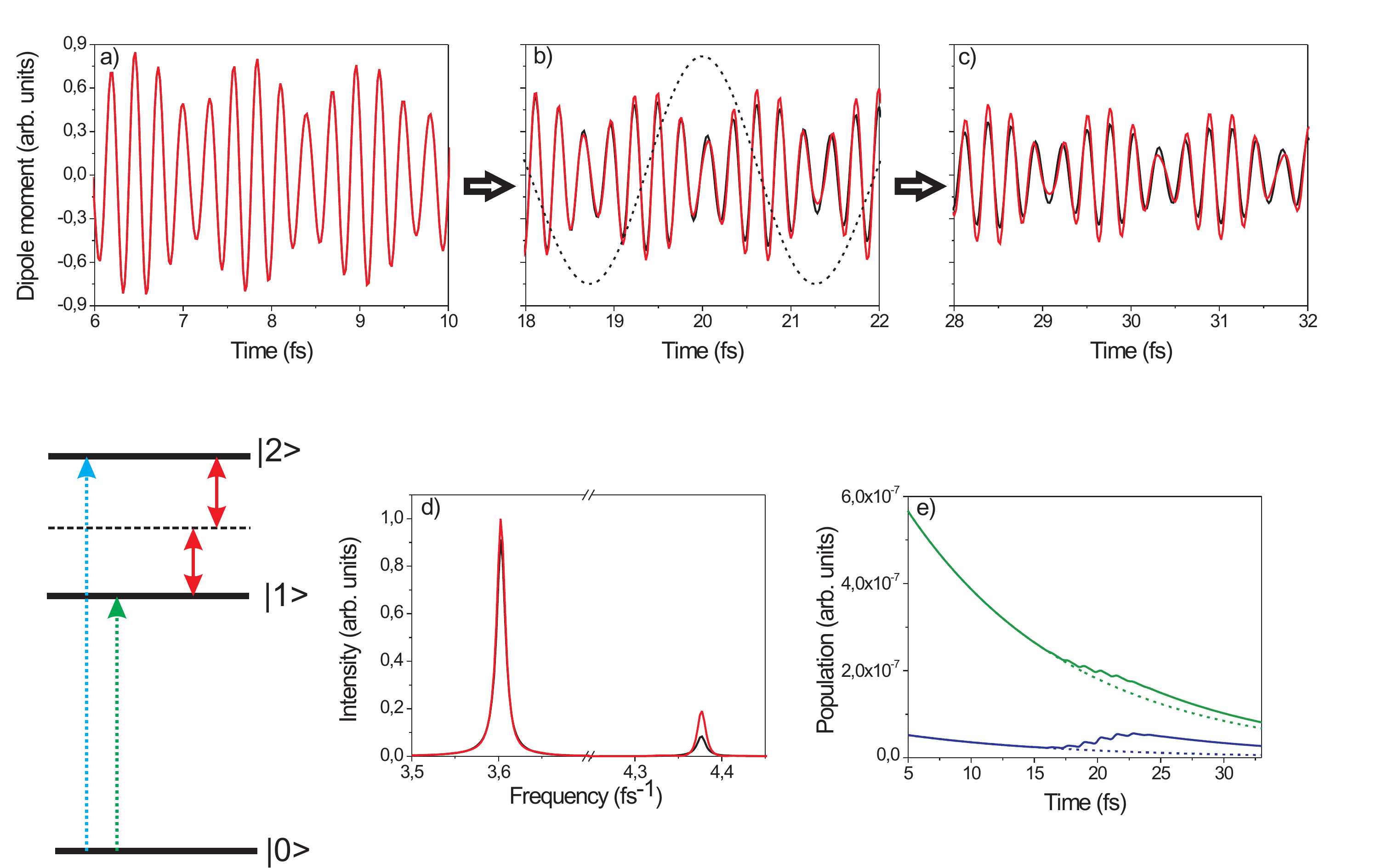}
\caption{Dipole oscillations for three time intervals [6-10 fs] (a), [18-22 fs](b), and [28-32 fs](c) without (black line) and with IR field (red line). The pulse (black dotted line) arrives at a delay $t$=20 fs, with respect to the initial XUV excitation.(d) Fourier transform of the dipole moment without (black line) and with (red line) IR field .(e) Population of the upper (blue) and lower (green) excited states without (dashed lines) and with (solid lines) IR field. In the simulation a two-photon coupling between the excited states was considered.}
\label{Fig1}
\end{figure}

In order to illustrate the excitation of the electronic wave packets and observation of the time delay dependent patterns in the transmitted spectra, we consider a three level system interacting with a broadband XUV pulse and a synchronized CEP-stable IR field. In spite of its simplicity, the model allows one to gain a simple, intuitive picture of the experimental results and simulations, pointing out the generality of the approach. The population, initially in the ground state $\psi_0$, is raised (at the instant $t=0$) to two excited states $\psi_1$ and $\psi_2$ by interaction with an attosecond XUV pulse creating a coherent wave packet (see Fig. \ref{Fig1}):
\begin{equation}
\psi(t)=c_2(t)e^{[-i(E_2/\hbar-i\Gamma_2/2)t]}\psi_2+c_1(t)e^{[-i(E_1/\hbar-i\Gamma_1/2)t]}\psi_1+c_0(t)e^{[-iE_0t/\hbar]}\psi_0
\label{Eq1}
\end{equation}
where $c_0(t)$, $c_1(t)$, and $c_2(t)$ are the time-dependent amplitudes of the ground, the first, and second excited state, characterized by energy $E_0$, $E_1$, and $E_2$, respectively. We assume that the first and second excited states decay with time constants $1/\Gamma_1$ and $1/\Gamma_2$, respectively.

The attosecond pulse sets in a dipole oscillation $d(t)$, which is given by the sum of two contributions related to the coherent evolution of the states $\psi_0,\psi_1$ and $\psi_0,\psi_2$ (assuming no direct coupling between the two excited states): $d(t)=x_1(t)+x_2(t)$. The oscillation in time of the dipole moment (shown in Fig.~\ref{Fig1}a,b,c) is a direct manifestation of the coherent evolution of the molecular electronic charge associated to the three states.
The fast modulations are due to the energy differences $E_1-E_0$ and $E_2-E_0$, while the slow variation of the amplitude is connected to the beating of the two dipoles with energy difference $E_2-E_1$. The microscopic dipole moment determines a macroscopic polarization of the medium that affects both the phase (dispersion) and intensity (absorption/emission) of the XUV pulse transmitted through the medium.
Information about the dynamical evolution of the system are, therefore, encoded in the spectral phase and amplitude of this radiation.

In the temporal domain, a perturbation of the freely-evolving dipoles, such as the interaction with a moderately intense IR pulse, will influence the evolution of the complex coefficients $c_1(t)$ and $c_2(t)$, thus leading to modifications in the oscillating dipoles. In Fig.~\ref{Fig1} we consider (in the framework of the time-dependent perturbation theory) the interaction of the three level system with a few-cycle IR pulse arriving at a delay $t=20$~fs after the initial excitation. Before the arrival of the IR pulse, the dipole oscillation with (red line) and without (black line) IR laser field are indistinguishable, as shown in Fig.~\ref{Fig1}a. Around the delay $t=20$~fs (Fig.~\ref{Fig1}b), the IR field (black dotted line) modifies the oscillations with respect to the unperturbed case.
In terms of oscillating dipoles, the action of the IR field will depend on the relative phase between the electric field oscillations and the beating between the two dipoles at frequency $(E_2-E_0)/\hbar$ and $(E_1-E_0)/\hbar$. The modifications of the dipole oscillations remain encoded after the end of the IR pulse as shown in Fig.~\ref{Fig1}c.

In the spectral domain, the perturbation induced by the IR field leads to variations of the radiation transmitted through the sample (enhanced or reduced absorption or emission, and dispersion of the XUV light) as shown in Fig.~\ref{Fig1}d. The amplitudes $c_1(t)$ and $c_2(t)$ and also the populations of the excited states are modified sub-cycle by the IR field as shown in Fig.~\ref{Fig1}e. Due to the coherence between the ground state and the excited states the dipole oscillations can then be used as a reference to monitor the coupling between the states $\psi_1$ and $\psi_2$ induced by the IR field.

\begin{figure}[htb]
\centering\includegraphics[width=8cm]{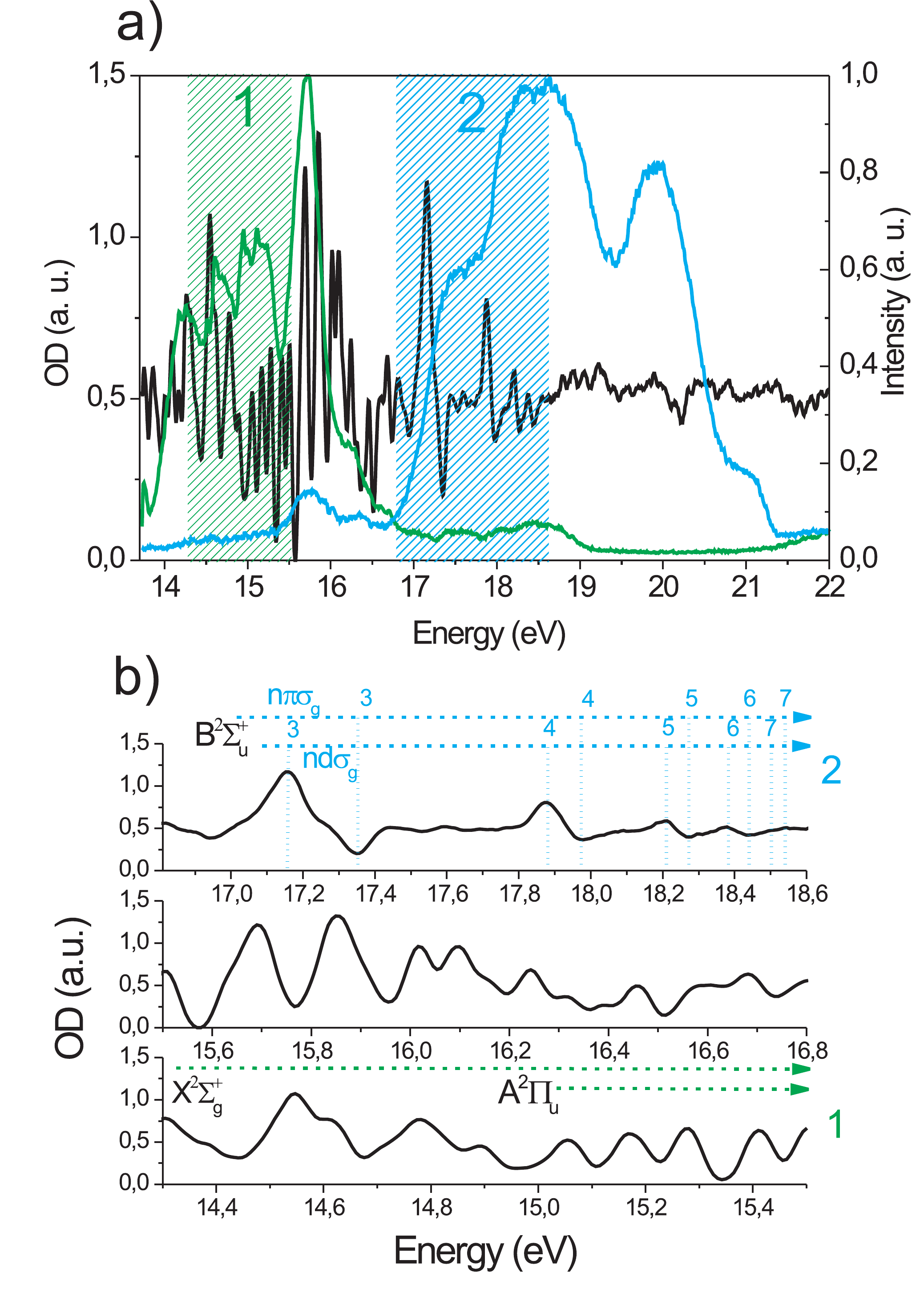}
\caption{a) Nitrogen optical density in the XUV spectral region (left axis, black line) measured using the XUV radiation generated by high-harmonic radiation in xenon. Normalized spectral intensity of the XUV radiation generated in xenon and transmitted through a 150-nm-thick indium (green curve) or tin (blue curve) filter (right axis). The relevant energy ranges are indicated by green (bound states) and blue (Fano resonances) lines.
b) Enlarged view of the measured nitrogen optical density. The blue lines (upper panel) indicate the position of the levels, characterized by quantum number $n$, belonging to the two series of Fano resonances (Hopfield absorption and apparent emission). The green lines (lower panel) indicate the energy region of features corresponding to excited levels converging to the ground ($X^2\Sigma_g^+$) and first excited state ($A^2\Pi_u$) of $\mathrm{N_2^+}$, respectively.}
\label{Fig2}
\end{figure}
In spite of its simplicity the three-state model allows one to describe several general aspects of the physical mechanism occurring in the interaction of an XUV pulse and a synchronized IR field with simple (atoms) and more complex systems (molecules).\\

In our experiment, we have investigated the electronic dynamics occurring in nitrogen after the creation of a broadband wave packet using different attosecond pulses.
Differently from the simple three-state model, the cross-section of nitrogen in the vacuum ultraviolet region (14-22 eV) is characterized by several features corresponding to the excitation of the neutral molecule and of the molecular ion, as shown in Figs.~\ref{Fig2}a,b.

In order to quantify the absorption the nitrogen sample, we introduce the optical density defined as:
\begin{equation}
\mathrm{OD}(\omega)=-\ln\Big[\frac{S(\omega)}{S_0}\Big]
\end{equation}
where $S(\omega)$ is the XUV spectrum transmitted through the sample and $S_0(\omega)$ is the reference spectrum acquired without nitrogen.
The radiation transmitted through the 3-mm-thick gas cell was analyzed using an XUV spectrometer composed of a toroidal mirror and a concave
grating, which disperses the XUV light in the focal plane where a MCP coupled to a phosphor screen was placed. A CCD camera acquired the signal at the back of the phosphor screen. The design of the spectrometer ensures a high spectral resolution in the range 15-25 eV ($\Delta E=30$~meV) that allows us to partially resolve the rich level structure of nitrogen in the 14-22~eV energy range.

The features visible in Fig.~\ref{Fig2}a are attributed to the Rydberg series converging to the ground ionic state ($X^2\Sigma_g^+;I_p=15.58~\mathrm{eV}$) and to the first ($A^2\Pi_u;E=16.94~\mathrm{eV}$) and second ($B^2\Sigma_u^+;E=18.75~\mathrm{eV}$) ionic excited states.

The energy region 16.8-18.6 eV (Fig.~\ref{Fig2}b upper panel) is characterized by features corresponding to the Hopfield absorption and apparent emission series~\cite{PR-Hopfield-1930}. These series correspond to two series of Fano(-Beutler) autoionizing
resonances converging to the $B^2\Sigma_u^+$ excited ionic state. The Hopfield absorption and emission series have been attributed to the excitation of a $2\sigma_u$ electron to states belonging to the Rydberg series of $\mathrm{N_2}$ $nd\sigma_g~^1\Sigma_u^+$ and
$nd\pi_g~^1\Pi_u$, respectively~\cite{PS-Berg-1991,JPB-Raoult-1983}. Due to the coupling with the continuum, autoionization leads to the
emission of an electron leaving the ion either in the $X^2\Sigma_g^+$ or in the $A^2\Pi_u$ state~\cite{CPL-Gurtler-1977}.

Between the ionization potential ($I_p=15.58$~eV) and 16.8~eV (Fig.~\ref{Fig2}b central panel),
the features are mainly attributed to two series of autoionizing resonances converging to the
$A^2\Pi_u$ state~\cite{PS-Berg-1991}.

Finally, below 15.6 eV (Fig.~\ref{Fig2}b lower panel) the absorption features are attributed to two series of excited states of the neutral molecule~\cite{CPL-Gurtler-1977, JCP-Huffmann-1963}, converging the ground ionic state $X^2\Sigma_g^+$, and to the first excited state $A^2\Pi_u$. Due to the finite energy resolution of the XUV spectrometer, the full structure of the excited states cannot be resolved.

The optical density was measured using broadband XUV continua generated in xenon by means of the polarization gating
technique~\cite{SCIENCE-Sansone-2006} and filtered using either indium or tin filters to obtain
spectra centered around 15~eV (green line) and 18~eV (blue line), respectively (see Fig.~\ref{Fig2}a).

As can be observed in Fig.~\ref{Fig2}a,b, the Fano-resonances (blue shaded area in Fig.~\ref{Fig2}a and upper panel of Fig.~\ref{Fig2}b) and the states below (around) the ionization threshold (green shaded area in Fig.~\ref{Fig2}a and lower panel of Fig.~\ref{Fig2}b) are spaced by roughly 3 eV, corresponding to the energy of two IR photons. According to the simple three-state model, the coherent excitation of these electronic states due to the absorption of a single XUV photon and the subsequent interaction with a synchronized IR field, should lead to a coupling of the dipole moment and to a modulation of the absorption optical density. In this picture the bound excited states and the Fano resonances correspond to the states $|1>$ and $|2>$ of the three-state model shown in Fig.~\ref{Fig1}.

The possibility to change the spectrum of the XUV continua using different metallic filters, allows one to modify the initial population of the electronic wave packet. In particular, using a tin filter, it is possible to generate an XUV continuum centered around 18~eV (blue spectrum in Fig.~\ref{Fig2}a), while using indium filter an XUV spectrum centered at 15~eV can be selected (green spectrum in Fig.~\ref{Fig2}a). Pulse durations of 927~as and 1.4~fs were retrieved using the FROG-CRAB algorithm for the two cases, respectively~\cite{PRA-Mairesse-2005}. According to the mean photon energy of the XUV pulses, we will refer hereafter to the dynamics initiated by the two XUV spectra as low-energy excitation case (LEE) and high-energy excitation case (HEE), respectively.
In the LEE case mainly the bound excited states of nitrogen are populated, whereas in the HEE case the two series of Fano resonances are
efficiently excited. In both cases, the tail of the XUV spectra [towards the high (low) energies for the LEE (HEE) cases] creates a broadband electronic wave packet with components below and
above the ionization threshold.

\begin{figure}[htb]
\centering\includegraphics[width=9.5cm]{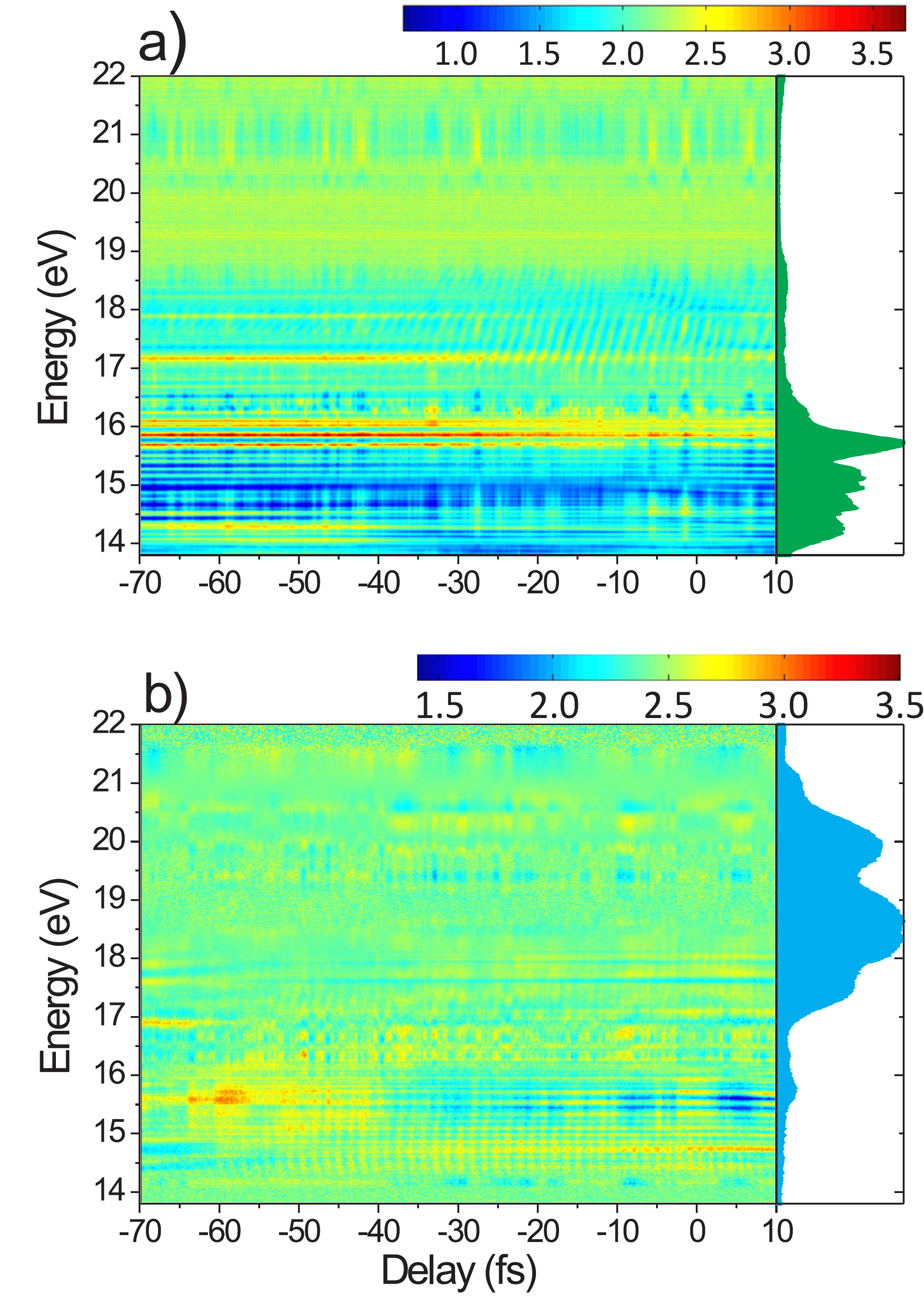}
\caption{Optical densities as a function of the relative time delay between the IR and XUV pulses for the low-energy excitation (LEE) case (a) and the high-energy excitation (HEE) case (b). The excitation XUV spectra are shown on the right hand-side. The XUV field comes first for negative delays.}
\label{Fig3}
\end{figure}

\begin{figure}[htb]
\centering\includegraphics[width=9.5cm]{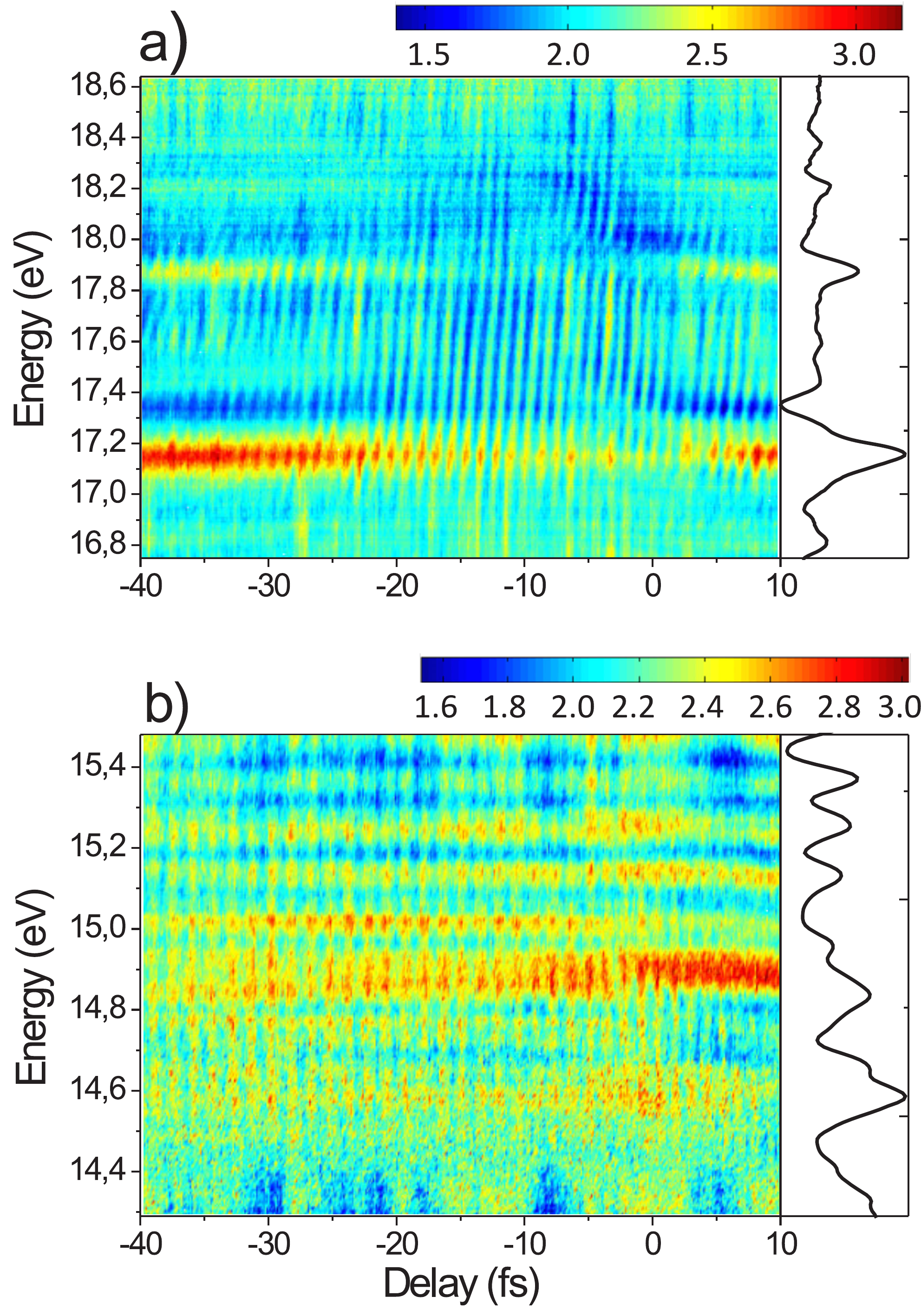}
\caption{Enlarged view of the the optical density as a function of the relative delay between the IR and XUV pulses for the low-energy excitation LEE case in the region of the Fano resonances (a) and for the HEE case in the region of the bound states (b). The optical densities measured without IR pulse are shown on the right hand-side as references.}
\label{Fig4}
\end{figure}

\begin{figure}[htb]
\centering\includegraphics[width=9.5cm]{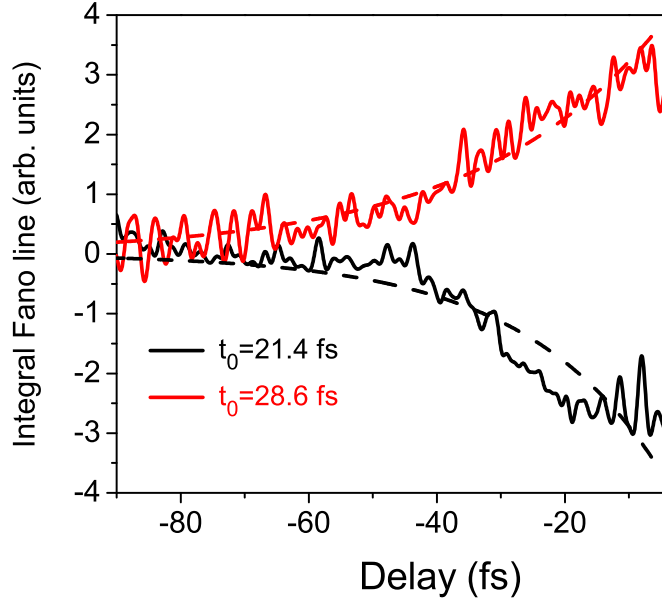}
\caption{Integrated cross-section signal (solid line) around the Fano resonances n=3 of the series $nd\sigma_g~^1\Sigma_u^+$ (black) and $nd\pi_g~^1\Pi_u$ (red) (see Fig.~\ref{Fig4}a) as a function of the relative delay between the IR and XUV pulses. The values indicate the time constants $t_0$ of the two exponential fits (dashed lines).}
\label{Fig5}
\end{figure}
\section{Sub-cycle quantum beating and autoionization dynamics}
In the time-resolved experiments, a moderately intense ($I\simeq10^{12}~\mathrm{W/cm^2}$) CEP-stabilized few-cycle IR pulse was
introduced, with a variable time delay, to overlap with the XUV continua in the nitrogen cell.
The optical density $\mathrm{OD}$ was measured as a function of the energy and of the relative time delay.
The IR dressing pulse arrives after the XUV pulse for negative delays.

The delay-dependence of the optical density is shown in Figures~\ref{Fig3}a~and~\ref{Fig3}b in the LEE and HEE cases with the corresponding XUV excitation spectrum (right hand-side), respectively. Clear oscillations with a period of $T=1.3$~fs (corresponding to half-optical cycle of the IR field) are observed in particular energy ranges in both cases. It is important to point out that in the LEE case, the oscillations are evident for the Fano resonances, i.e. in the energy region that is only weakly populated by the XUV radiation (see Figs.~\ref{Fig3}a~and~\ref{Fig4}a); similarly, in the HEE case
the oscillations are present only on the bound excited states (see Figs.~\ref{Fig3}b~and~\ref{Fig4}b). As already anticipated the energy distance between these two groups of states is about $3$~eV, indicating a two-photon coupling as origin of the observed modulation, in agreement with the simple three-state model described in the introduction.
In Fig.~\ref{Fig3}b, half-cycle oscillation extending down to -60~fs can be observed. The origin of these oscillations can be attributed to a low intensity pedestal of the synchronized IR field. Indeed, even though pulses as short as 5.0~fs were characterized using the FROG technique, the different dispersion compensation between the point where the pulses are characterized and the nitrogen cell can lead to a low intensity pulse pedestal affecting the main pulse.

We point out that half-cycle oscillations were observed also using CEP unstable driving pulses, and trains of attosecond pulses. This observation indicates that the key element for the observation of the modulation of the optical density is the synchronization between the attosecond pulses and the oscillation of the IR field. On the other hand, the possibility to (partially) shape and tune the XUV spectrum, allows one to enhance the modulation effect and to determine the spectral region where it is more evident.
Because of the spectral distribution of the XUV pulses (as explained in the following), such oscillations are not present on the autoionizing states converging to the $A^2\Pi_u$ state. These states will not be discussed further in this work. We have experimentally verified
that the contrast of the half-optical cycle oscillations strongly depends on the shape of the XUV spectra.
In particular, the maximum contrast was achieved when the tails of the XUV spectral distributions were about 5\%-10\% of the main peak.

Beside the half-optical cycle oscillation in the signal, an overall depletion in the autoionizing
region accompanied by broadening can be observed in Fig.~\ref{Fig4}a when the two pulses overlap and for short negative delays (the XUV field comes before the IR pulse), which indicates that the IR field ionizes these states before autoionization can occur. The level of
depletion at each autoionizing state decreases at negative delays at half the rate of the decay lifetime of that state, and the signal revives at large negative delays. This type of time-domain measurement of autoionization has been performed in previous
experiments~\cite{PRL-Wang-2010, PRL-Gilbertson-2010}.

The lifetime $\tau=1/\Gamma$ (where $\Gamma$ indicates the linewidth of the autoionizing photoelectron line) of the first two Fano resonances can be retrieved from the experimental data by analyzing the delay-dependence of the integral of the Fano profile. In order to do this, we have first eliminated the oscillation at frequency $2\omega$ by Fourier filtering and then integrated the optical density signal over the energy interval 17.0-17.25~eV and 17.25-17.45~eV corresponding to the $3d\sigma_g~^1\Sigma_u^+$ and $3d\pi_g~^1\Pi_u$ autoionizing states, respectively. The results are shown in Fig.~\ref{Fig5}. The exponential fits (dashed lines) indicate a time constant of $t_0=$21.4~fs and 28.6~fs, respectively. In the time domain, the decay dynamics of the integral signal is determined by the convolution of the cross-correlation of the excitation pulses (XUV and IR pulses) with the exponential decay of the polarization. The former one was estimated in $\simeq~8$~fs from the experimental rise (or decay) time of the signal at positive delays (see Fig.~\ref{Fig5}). We have verified that the decay constant of the polarization $\bar{t}$ and the time constant $t_0$ are approximatively given by $\bar{t}=0.93-0.97 t_0$, due to the non-zero cross-correlation. The lifetime $\tau$ of the autoionizing state is half the decay constant of the polarization $\bar{t}$~\cite{PRA-Bernhardt-2014}, leading to $\tau$=10.2~fs and 13.8~fs for the $3d\sigma_g~^1\Sigma_u^+$ and $3d\pi_g~^1\Pi_u$ resonances, respectively.

These values should be compared to the measurement of the linewidth reported in literature: according to G\"{u}rtler and coworkers~\cite{CPL-Gurtler-1977} the linewidth of the state n=5 of the series $nd\sigma_g~^1\Sigma_u^+$ is $\Gamma=$20~meV corresponding to a lifetime $\tau=32.9$~fs. Assuming a $n^3$~scaling (where $n$ indicates the principal quantum number) for the lifetime of the autoionizing states~\cite{JCP-Huber-1993}, a decay constant $\tau=7.1$~fs can be derived for the state $3d\sigma_g~^1\Sigma_u^+$ of the series.
In ref.~\cite{JCP-Huber-1993} the width of the state n=3 of the series $nd\sigma_g~^1\Sigma_u^+$ is about $\Gamma=$50.8~meV corresponding to a decay constant $\tau=12.9$~fs. We can therefore conclude that the measured decay time ($\tau$=10.2~fs) is intermediate and in reasonably good agreement with previously reported experimental results.


\section{Theoretical model and interpretation}
\begin{figure}[htb]
\centering\includegraphics[width=9.5cm]{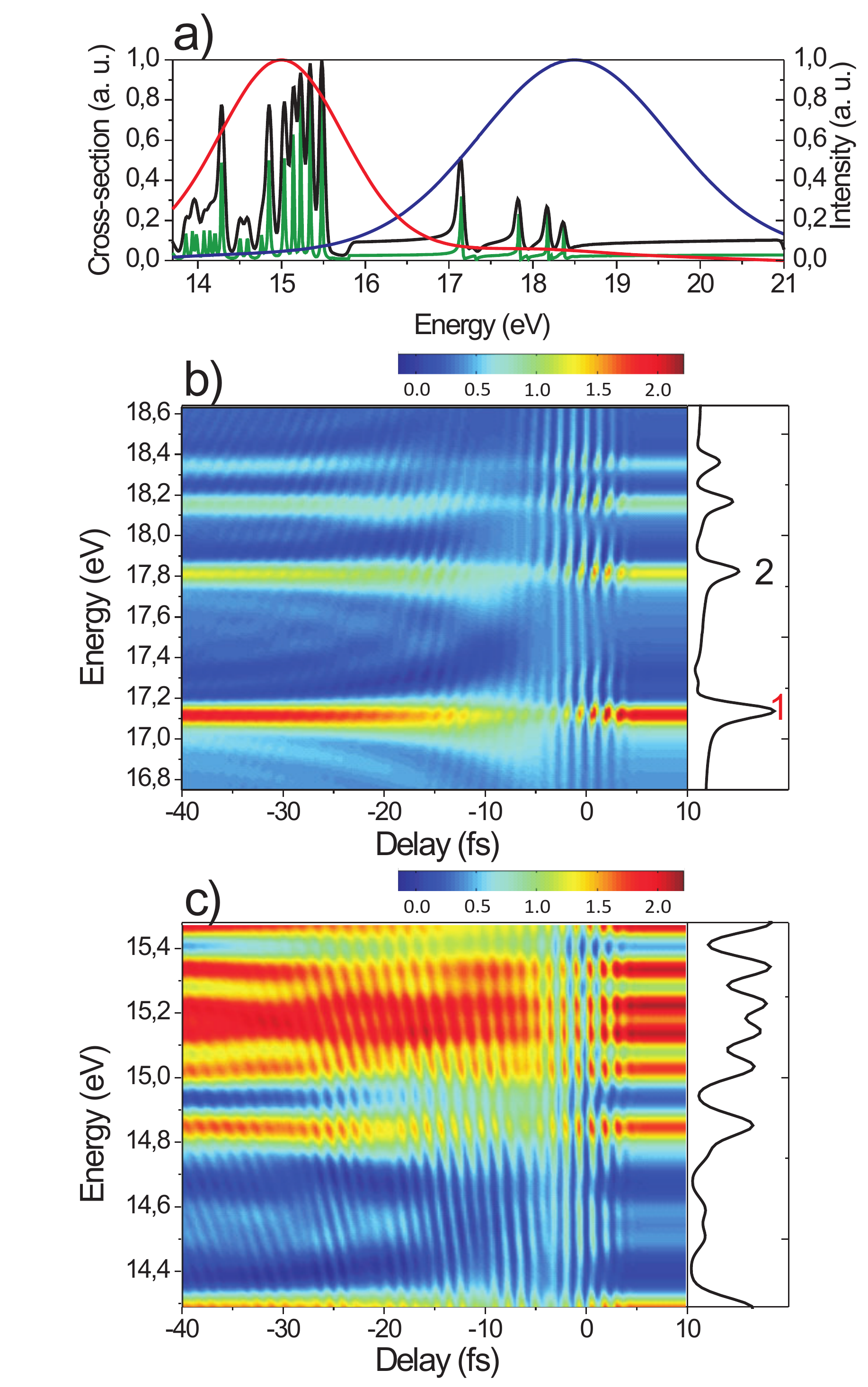}
\caption{a) Simulated cross-section in nitrogen using experimental data without (green line) and with convolution (black line) with the limited experimental spectral resolution (left axis). Simulated XUV spectral distributions for the LEE (red curve) and HEE (blue curve) cases. b) LEE: simulated cross-section as a function of the relative delay between the IR and XUV pulses in the spectral region of the Fano-Beutler resonances. The convoluted cross-section is shown on the right hand-side for reference. c) HEE: simulated cross-section as a function of the relative delay between the IR and XUV pulses in the spectral region corresponding to the bound excited states. The convoluted cross-section is shown on the right hand-side for reference. The two-dimensional cross-sections have been convoluted with the spectral response of the XUV spectrometer.}
\label{Fig6}
\end{figure}

The energy difference of about 3.1~eV between the two groups of levels suggests a two-photon
(single photon energy $\hbar\omega\approx 1.55$~eV) coupling between the Fano resonances
and the bound excited states, as origin of the half-optical cycle oscillations. We simulated
the experimental results using the time-dependent perturbation (TDP) theory, which is valid if all
transitions considered are weak in terms of transferred population. Two series of
Fano resonances and two series of bound excited states were considered in the simulation,
with the parameters extracted from the measured XUV spectra. To include IR ionization of the excited states, a decay rate
for each state as a function of the IR intensity was introduced. The rate was estimated with
general approximations regarding the coupling of autoionizing states by ultrashort pulses and the
associated light absorption~\cite{PRA-Chu-2013}.
The XUV pulse creates a broadband electronic wave packet composed by the ground
state $|g\rangle$, the bound states $|b_m\rangle$
below the ionization threshold $I_p$ and the autoionizing states $|f_n\rangle$ embedded in the background
continuum $|E\rangle$ above $I_p$. The general total wave function of the system is (atomic units are used unless otherwise specified)
\begin{equation}
|\Psi(t)\rangle = e^{-iE_gt} c_g(t) |g\rangle + e^{-i(E_g+\omega_X)t}
\left[ \sum_m {c_m(t) |b_m\rangle} + \sum_n {c_n(t) |f_n\rangle} +
\int {c_E(t) |E\rangle} \right]. \label{eq:wave}
\end{equation}
Here the fast oscillating terms are factored out where $E_g$ and $\omega_X$ are the ground state
energy and the XUV carrier frequency, respectively, and all the $c(t)$ coefficients of states near
the XUV energy are smooth in time. The indices $m$ and $n$ are used exclusively for the bound states and
Fano resonances, respectively.
The XUV spectral distributions mimic the experimental one, leading to a much more efficient population of the $|b_m\rangle$ or $|f_n\rangle$ states in the LEE and HEE case, respectively (see Fig.~\ref{Fig6}a). Meanwhile the IR pulse ($\lambda=780$~nm and FWHM=5~fs)
couples the two groups of states separated by two IR photon energies. The product of the IR intensity and the overall dipole matrix element is adjusted to maximize the oscillation contrast. Because of the significant imbalance of
populations, the IR mainly transfers the electrons from the highly populated group to the weakly
occupied group, i.e., either from $|b_m\rangle$ to $|f_n\rangle$ or from $|f_n\rangle$ to
$|b_m\rangle$. In the LEE case, the final $|f_n\rangle$ states, while weakly populated, consist
of the electrons from the direct XUV excitation and from the XUV+2IR three-photon excitation
through the $|b_m\rangle$ states. Thus, the spectrum around $|f_n\rangle$ shows the interference
pattern formed by the two paths. Similarly, in the HEE case, where XUV concentrates on
$|f_n\rangle$, the interference pattern is observed around $|b_m\rangle$.
In this model the IR transition brings just equivalently small amount of population to the weakly
pumped states. The XUV photon energy is assumed to be much higher than its bandwidth. We apply
rotating wave approximation and slowly varying envelope and write the XUV electric field in the
general form of
\begin{equation}
E_X(t) = F_X(t) e^{i\omega_Xt} + F_X^*(t) e^{-i\omega_Xt}, \label{eq:FX}
\end{equation}
where $|2F_X(t)|^2$ is the intensity profile of the pulse, and $F_X(t)$ contains any phase
information additional to the carrier frequency. Focusing on the LEE case first, taking the time
derivative form of TDP, the coefficients of the autoionizing states $|f_n\rangle$ are calculated by
\begin{equation}
\dot{c}_n(t) = i\bar{D}_{ng} F_X^*(t) - \sum_m {D_{nm} E_L^2(t) c_m^{(0)}(t)} + \left[ i\delta_n -
\kappa_n - w_n(t) \right] c_n(t). \label{eq:cn}
\end{equation}
The first term is the contribution by the XUV transition from the ground state, where the ground
state coefficient is assumed to remain 1. The complex dipole matrix elements $\bar{D}_{ng} \equiv
D_{ng}(1-i/q_n)$ are between $|g\rangle$ and the Fano resonances at $|f_n\rangle$, which contain
the Fano $q$ parameters and encode the Fano line shapes in the coefficients $c_n(t)$.
The second term is the contribution by the two-photon IR transition from all the $|b_m\rangle$
states, where the dipole matrix elements $D_{nm}$ represent the two-photon transition matrix
summing over all intermediate states, assuming none of them is resonant~\cite{PRA-Meshulach-1999}.
The coefficients $c_m^{(0)}(t)$ are prepared by solving
\begin{equation}
\dot{c}_m^{(0)}(t) = iD_{mg} F_X^*(t) + \left[ i\delta_m - \kappa_m - w_m(t) \right] c_m^{(0)}(t).
\label{eq:cm0}
\end{equation}
In the last term on the right hand side of Eq.~(\ref{eq:cn}), $\delta_n \equiv E_g + \omega_X -
E_n$ is the detuning of the XUV with respect to state $|f_n\rangle$, and $\Gamma_n \equiv
2\kappa_n \equiv 2\pi V_n^2$ and $2w_n(t)$ are the resonance width and IR ionization rate of
$|f_n\rangle$. Note that the $\delta_m$ and $\delta_n$ in Eqs.~(\ref{eq:cn}) and (\ref{eq:cm0}) do not appear in the
common TDP formulas written in Dirac picture because the phase terms of the excited states in our
total wave function are defined relative to the XUV central frequency but not to the state energy
levels. The IR ionization rates $w_n(t)$ and $w_m(t)$ are assumed to be proportional to the IR
intensity because these states are close to the ionization threshold in terms of the IR photon
energy, and single photon or multiphoton ionization instead of tunnel ionization more suitably
describes the process. The actual rates are estimated by fitting with the present experimental
data. For the HEE case, similar procedure described above is taken. The coefficients
of $|b_m\rangle$ are calculated by
\begin{equation}
\dot{c}_m(t) = iD_{mg} F_X^*(t) - \sum_n {D_{mn} E_L^2(t) c_n^{(0)}(t)} + \left[ i\delta_m -
\kappa_m - w_m(t) \right] c_m(t), \label{eq:cm}
\end{equation}
where $c_n^{(0)}(t)$ are given by
\begin{equation}
\dot{c}_n^{(0)}(t) = i\bar{D}_{ng} F_X^*(t) + \left[ i\delta_n - \kappa_n - w_n(t) \right]
c_n^{(0)}(t). \label{eq:cn0}
\end{equation}
Here we obtain the total wave function beside the background continuum for any given set of XUV
and IR pulses and parameters. The exact calculation for the continuum part is not
necessary for the absorption spectrum as will be shown later.

It is to be emphasized that all optical transitions in the current model are based on TDP, so each
transition is weak and has only one direction, from the highly populated state (as the initial
state) to the weakly populated state (as the final state). The conservation of electronic probability
(or of energy) does not apply because the initial state is not changed by the transition. The
strong coupling involving Rabi flopping or involving dressed states is excluded in this model.
Thus, many features observed in the strongly coupled transient absorption
experiments~\cite{CP-Loh-2008, NATURE-Ott-2014} cannot be recovered here.

To determine the input atomic parameters, the experimental data were analyzed. The line
widths of the lowest two autoionizing states at 17.15 and 17.32~eV are extracted from the signal
along the resonances in the time-delayed XUV absorption measurement (the spectral resolution of the XUV spectrometer was taken into account).
The $q$ parameters of the same two states are extracted from the field-free XUV absorption spectroscopy. These two
states are the most easily observed states belonging to the two series that we consider.
The widths of higher resonances can be estimated by the quantum defect rule $\Gamma \nu^3 = const.$ for each series~\cite{JCP-Huber-1993}, where $\nu$ are the effective quantum numbers, which can be approximated
by the principal quantum numbers $n$ for low angular momentum. The $q$ parameter for each series is
assumed to be constant. The parameters for the autoionizing states are listed in
Table~\ref{tab:auto}.
\begin{table}[htbp]
\centering
\caption{Parameters of the autoionizing states used in the main simulation.}
\begin{tabular}{c c c c}
\hline\hline
$n$ & $E_r$(eV) & $\Gamma$(meV) & $q$\\
\hline
3 & 17.15 & 30.8 & -3.0 \\
3 & 17.32 & 23.0 & -0.5 \\
4 & 17.84 & 13.0 & -3.0 \\
4 & 17.94 & 9.7 & -0.5 \\
5 & 18.18 & 6.7 & -3.0 \\
5 & 18.23 & 5.0 & -0.5 \\
6 & 18.37 & 3.9 & -3.0 \\
\hline
\end{tabular}
\label{tab:auto}
\end{table}
The energy levels and line widths of the bound states $2\kappa_m$ are extracted from the
field-free XUV absorption measurement and listed in Table~\ref{tab:bound}.
\begin{table}[htbp]
\centering
\caption{Parameters of the bound excited states used in the main simulation. The line shapes
are assumed Lorentzian.}
\begin{tabular}{c c c c}
\hline\hline
$E_r$(eV) & $\Gamma$(meV) \\
\hline
13.98 & 0.02 \\
14.07 & 0.02 \\
14.14 & 0.02 \\
14.22 & 0.02 \\
14.28 & 0.02 \\
14.50 & 0.02 \\
14.59 & 0.02 \\
14.72 & 0.02 \\
14.85 & 0.02 \\
15.03 & 0.02 \\
15.14 & 0.02 \\
15.23 & 0.02 \\
15.36 & 0.02 \\
15.48 & 0.02 \\
\hline
\end{tabular}
\label{tab:bound}
\end{table}
The relative strengths of the dipole matrix elements between the bound excited states and the
autoionizing states are assumed to be proportional to $V_m V_n$. The overall strength between the
group of bound excited states and the group of autoionizing states is adjusted to maximize the
contrast of the interference fringes in the main result.
The light absorption of the system corresponds to its dipole oscillation in time $D(t) = \langle
\Psi(t) |D| \Psi(t) \rangle$. The XUV absorption thus corresponds to the oscillation in the XUV
frequency range, which for the total wave function in Eq.~(\ref{eq:wave}) can be reduced
to~\cite{PRA-Chu-2013}
\begin{equation}
D(t) = e^{i\omega_Xt} \left[ \sum_m \bar{D}_{gm} c_m^*(t) + \sum_n D_{gn} c_n^*(t)
- i \pi F_X(t) |D_{gE}|^2 \right] + c.c., \label{eq:D}
\end{equation}
where we have assumed $c_g(t)=1$ and taken the approximation where $D_{gE}$ is constant of $E$
estimated at the lowest $f_m$. Note that for each Fano resonance, a given set of constant
$D_{gE}$, resonance width $\Gamma_n$, and $q_n \equiv D_{gn}/(\pi V_n D_{gE})$ determines the
dipole matrix element $D_{gn}$. Further details for autoionizing systems coupled by an ultrashort
intense laser have been discussed in Ref.~\cite{PRA-Chu-2013}. The present observed experimental
features are from the interference between the two paths introduced above, where only their
relative but not absolute strengths are essential, and each path is formed by the dipole matrix
elements coupled with the field strength. The absorption of the XUV light is obtained through the
response function~\cite{PRA-Gaarde-2011} given by
\begin{equation}
S(\omega) = -2 \mathrm{\mathrm{Im} \left[ \tilde{D}(\omega) \tilde{E}_X^*(\omega) \right]}, \label{eq:S}
\end{equation}
which represents the absorption probability density in frequency, where the convention for
Fourier transform is
\begin{equation}
\tilde{f}(\omega) = \frac{1}{\sqrt{2\pi}} \int{ e^{-i\omega t} f(t) dt}. \label{eq:ft}
\end{equation}
The absorption cross section $\sigma(\omega)$ in the XUV energy range, which is proportional to the optical density $\mathrm{OD}(\omega)$ reported in the experiment, is related to the response function by
\begin{equation}
\sigma(\omega) = \frac{4\pi \alpha \omega S(\omega)}{|\tilde{E}_X(\omega)|^2}, \label{eq:xsec}
\end{equation}
where $\alpha$ is the fine structure constant.

The simulated cross section is reported in Fig.~\ref{Fig6}a (green line) with the corresponding XUV spectra of the attosecond pulses used in
the simulations. In order to compare the outcome of the simulations with the experimental data,
the resolution of the XUV spectrometer was included, by convoluting the simulated cross section
with a gaussian function, whose width corresponds to the spectral resolution (black line in
Fig.~\ref{Fig6}a). The simulations, shown in Fig.~\ref{Fig6}b and \ref{Fig6}c, clearly reproduce the periodic
oscillations and the IR-induced depletion and broadening of the time-delayed signals in both the
LEE and the HEE cases, in agreement with the experimental results. The presence of these
oscillations supports the conclusion of an IR-coupling occurring between the two groups of states. The
slopes of the oscillation fringes also fit the experiment well. The simulations allow to conclude that in the LEE (HEE) case,
when the XUV light pumps effectively the bound states (Fano resonances), the amplitudes of the Fano resonances (bound
states) are the result of the coherent sum of the weak direct XUV-excitation path (due to the tail of the XUV spectrum) and the XUV+2IR photons excitation mechanism through the bound states (Fano resonances).
\begin{figure}[htb]
\centering\includegraphics[width=8cm]{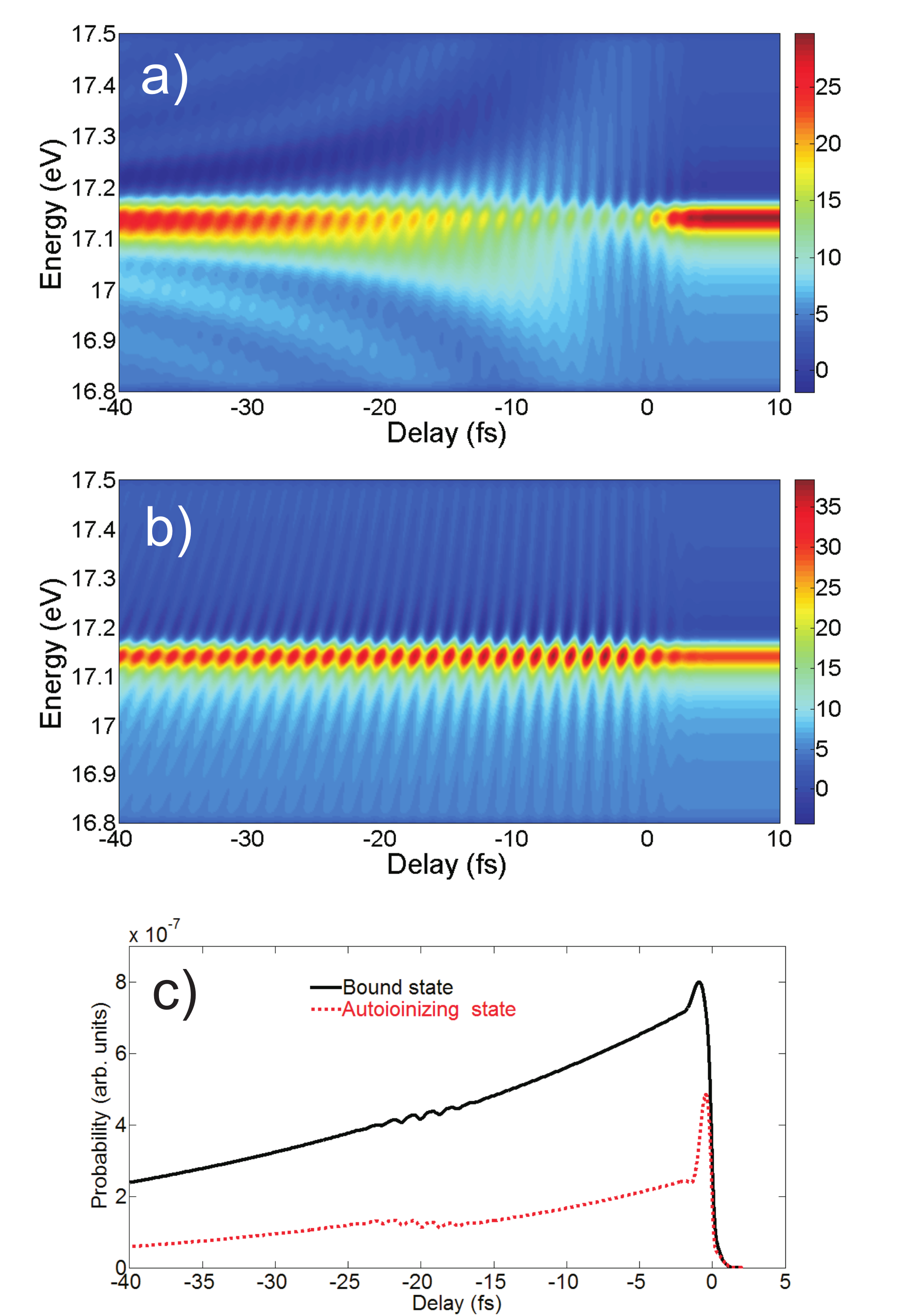}
\caption{Simulated delay-dependent cross section near the lowest Fano resonance in the three-level
system (ground state, bound state at 13.98~eV, and Fano resonance at 17.15~eV.) in the LEE case,
(a) with and (b) without the ionization by the IR pulse. (c) The
probabilities of the bound state and the autoionizing state evolving in time where the XUV peak
is at $t=0$~fs and the IR peak is at $t=-20$~fs, without the IR ionization.}
\label{Fig7}
\end{figure}

In order to support this main conclusion, we conducted a simulation
with the same field parameters of the LEE case but in a reduced system composed by only three
quantum levels--the ground state, the bound excited state at 13.98~eV, and the Fano resonance at
17.15~eV. The two excited levels are separated energetically by two IR photons. As seen in
Fig.~\ref{Fig7}a, the half-optical-cycle oscillation and the time-dependent depletion are reproduced very
well. 
It confirms that the main features shown in the experiment are already
manifested by the coupling in a simple three-level system. Nonetheless, there still exist some
noticeable differences from the full simulation. For instance, the spectral span near the zero time
delay is narrower, and the nearly vertical fringes shrink to only short pieces at the resonance.
Furthermore, the overall signal strength is weaker, which brings down the sharpness and the
contrast of the interference pattern as well. It shows that the interference effect in our
experiment is the collective contribution by all the direct (XUV) and indirect (XUV+2IR)
excitations.

When we further remove the IR ionization (Fig.~\ref{Fig7}b), the shift of the slope of the
half-optical-cycle fringes becomes very regular along the time delay, which has been demonstrated
in photoelectron spectroscopy theoretically~\cite{PRA-Choi-2010} and
experimentally~\cite{PRL-Kim-2012}. In the time domain, Fig.~\ref{Fig7}c shows the populations of the two
excited states, where the peak of IR is 20~fs later than the peak of XUV (0~fs). After being pumped by
the XUV, both excited states drop exponentially with their respective decay lifetimes. At t=-20~fs, the interaction with the
IR pulse redistributes the components of the wave packet, and the autoionizing state
receives a jump in the population. These simulations confirm the simple physical picture outlined in section 2.\\

\begin{figure}
\centering \resizebox{0.9\hsize}{!}{\includegraphics{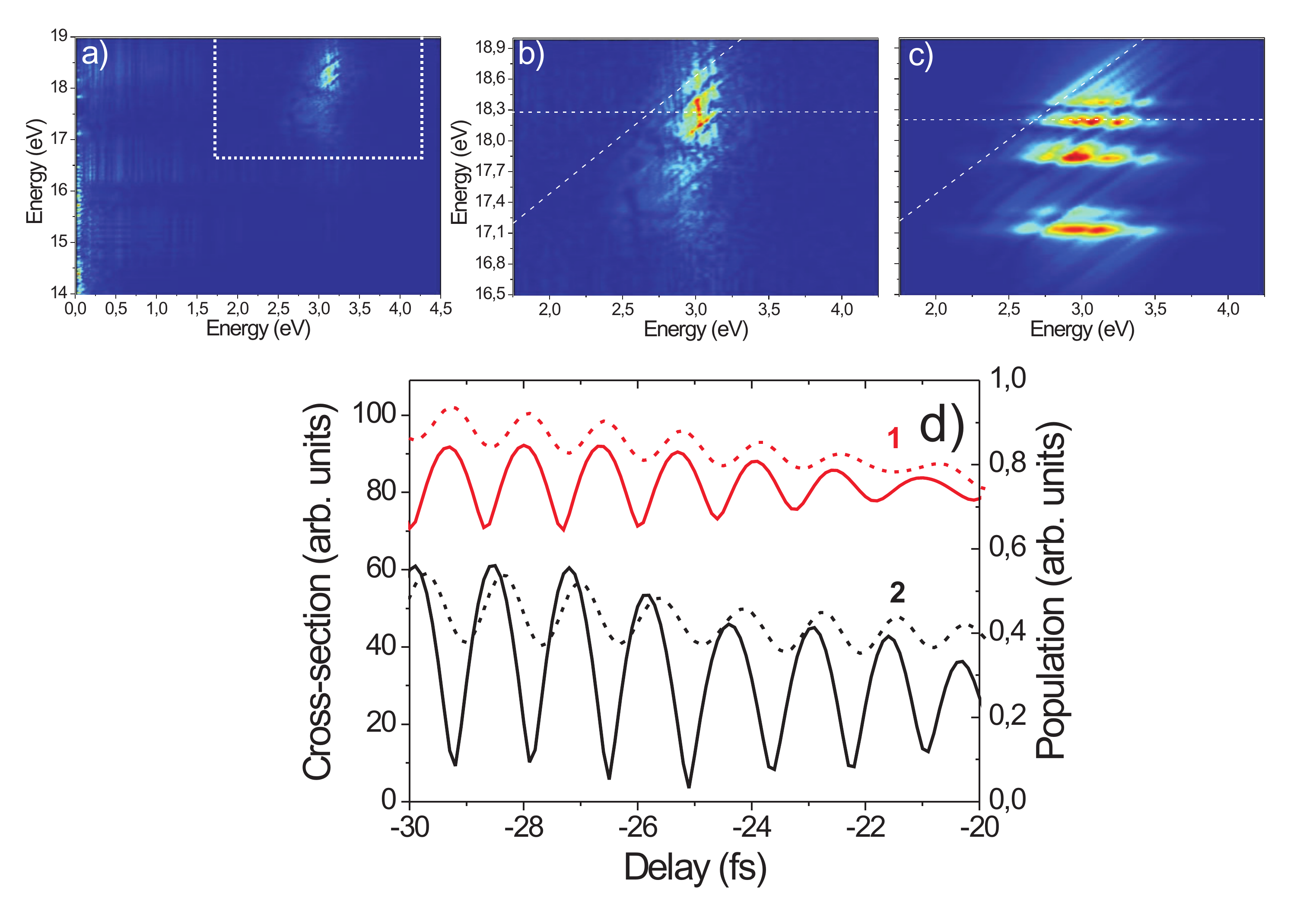}}
\caption{(a) Fourier transform of the delay-dependent cross-section shown in Fig.~\ref{Fig3}b (LEE case). (b) Enlarged view of the region around $\omega=17.5$ eV and $\omega'=3.1$ eV delimited by dashed lines in (a). (c) Fourier analysis of the simulated delay-dependent cross-section shown in Fig.~\ref{Fig6}b. The dotted white lines are guides for the eyes. (d) Simulated delay-dependence of the final population (solid lines) and of the cross-section (dashed lines) for the Fano-resonance with energy 17.1~eV (1; red), and with energy 17.8~eV (2; black), shown in Fig.~\ref{Fig6}b. The curves have been vertically displaced for visual clarity.}
\label{Fig8}
\end{figure}

The energy position of the interference fringes is determined by the initial population of the electronic wave packet components as shown in Figs.~\ref{Fig8}a, which reports the Fourier
transform along the delay axis of the energy-delay scan for the LEE case (Fig.~\ref{Fig3}b).
The half-cycle oscillations as a function of the delay correspond
to the features visible at energies $\omega'=2\omega_{IR}$ ($\approx 3.1$ eV) around the
Fano resonances (17.0-18.6~eV). The situation reverses for the
HEE case, where features around $\omega'=2\omega_{IR}$ are visible around the energies of the bound
states (14.5-15~eV) (not shown). In both cases, the group of states that is more
effectively populated by the XUV pulse present a much reduced oscillation with respect to the other one.
This observation points out that the cross-section modulations can be observed only if the two interfering paths present similar contributions to the total cross-section. For the group of state directly populated by single-XUV-photon absorption, the second path, characterized by XUV+2IR photons, is too weak too produce a clear interference and a measurable cross-section oscillation.
Therefore, the spectrum of the attosecond pulse determines which group of states will present the most pronounced absorption modulation due to the two-photon coupling.
A closer look at the LEE case (Fig.~\ref{Fig8}b) reveals that the signal at
$2\omega_{IR}$ is characterized by a series of tilted lines oriented at 45 degree in the
$(\omega,\omega')$ plane. The Fourier analysis of the simulations (see Fig.~\ref{Fig8}c) shows a
similar qualitative behavior confirming the presence of these lines around the position
$\omega'=2\omega_{IR}$. Each of these
feature is associated to a XUV+2IR path through a single bound excited state. Exploiting the
beating between different components of the electronic wave packet, we can dynamically
enhance with an attosecond time resolution the population of the Fano resonances in the LEE case and of
the bound excited states in the HEE case. This control is directly mapped in the delay dependence of the cross-section as shown in Fig.~\ref{Fig8}d, which reports the final population (i.e. after the interaction with both pulses) of the two Fano resonances (states
indicated by 1 and 2 in Fig.~\ref{Fig6}b) and the corresponding cross-section as a function of the
relative delay between the XUV and IR pulses. The relative delays between -30~to -20~fs
ensure that the IR control field arrives well after the initial excitation of the XUV
pulse. The oscillation of the population of the excited states are directly mapped on the
modulation of the transmitted XUV spectrum, thus allowing a straightforward visualization of the
variations of the electronic wave packet induced by the IR field.
\section{Conclusions}
In conclusion, we have investigated the ultrafast electronic dynamics occurring in a nitrogen
molecule excited by a tunable XUV continuum. The decay time of two Fano resonances can be retrieved exploiting the ionization induced by a few-cycle IR field.
At the same time, the IR pulse perturbs the coherent dipole oscillations inducing modulations in the transmitted XUV radiation, which are sensitive to the relative amplitude and relative phase of the components of the electronic wave packet. The
energy range where such modulations are enhanced can be controlled by manipulating the initial
state population. The use of tunable IR/visible fields or temporally shaped pulses might allow the investigation of the spectrally and temporally
resolved beating of selected components of a broadband wave packet. By combining spectral and
phase characterization of the transmitted XUV field, it will be possible to infer additional
information about the time evolution of the electronic wave packet with the aid of theoretical
modeling.
\section{Acknowledgments}
Financial support by the Alexander von Humboldt Foundation (Project ``Tirinto''), the
Italian Ministry of Research (Project FIRB No. RBID08CRXK),the European Research Council
under the European Community's Seventh Framework Programme (FP7/2007-2013) / ERC grant
agreement n. 227355 - ELYCHE, the MC-RTN ATTOFEL (FP7-238362)ATTOFEL, LASERLAB-EUROPE (grant agreement n. 284464, EC's Seventh Framework Programme), and Chemical Sciences, Geosciences and Biosciences
Division, Office of Basic Energy Sciences, Office of Science, U. S. Department of Energy,
is gratefully acknowledged. This project has received also funding from the European Union's Horizon 2020 research and innovation programme under the Marie Sklodowska-Curie grant agreement no.~641789 MEDEA. The authors also want to thank C. Ott for fruitful
discussions.

\section*{References}
\bibliography{NJP_bibliography}

%

\end{document}